\documentclass[twocolumn,prd,groupedaddress,nofootinbib]{revtex4}

\usepackage{amsfonts,amsmath,amssymb,mathrsfs}
\usepackage{hyperref}
\usepackage{color}
\usepackage{graphicx}  
\usepackage{dcolumn}   
\usepackage{bm}        
\usepackage[english]{babel}
\def\a{\alpha}

\def\r{\rho}
\def\s{\sigma}
\def\t{\tau}
\def\m{\mu}
\def\n{\nu}
\def\k{\kappa}
\def\th{\theta}
\def\g{\gamma}\def\G{\Gamma}
\def\L{\Lambda}\def\l{V}
\def\D{\Delta}
\def\la{\langle}
\def\ra{\rangle}
\def\o{\omega}\def\O{\Omega}
\def\d{\delta}
\def\p{\partial}

\def\oxthree{{\cal O}(x^3) }

\def\half{\textstyle{\frac{1}{2}}}

\def\bdoc{\begin{document}}
\def\edoc{\end{document}}
\def\bea{\begin{equation}}
\def\eea{\end{equation}}

\def\beq{\begin{eqnarray}}
\def\eeq{\end{eqnarray}}
\def\ben{\begin{enumerate}}
\def\een{\end{enumerate}}
\def\la{\langle}
\def\ra{\rangle}
\def\a{\alpha}
\def\g{\gamma}\def\G{\Gamma}
\def\d{\delta}\def\D{\Delta}
\def\e{\epsilon}
\def\z{\zeta}

\def\th{\theta}
\def\k{\kappa}
\def\l{\lambda}
\def\m{\mu}
\def\n{\nu}
\def\o{\omega}
\def\p{\pi}
\def\r{\rho}
\def\s{\sigma}
\def\t{\tau}
\def\L{{\cal L}}
\def\S{\Sigma }
\def\gsim{\; \raisebox{-.8ex}{$\stackrel{\textstyle >}{\sim}$}\;}
\def\lsim{\; \raisebox{-.8ex}{$\stackrel{\textstyle <}{\sim}$}\;}
\def\gtrsim{\gsim}
\def\lessim{\lsim}
\def\loc{{\rm local}}
\def\vm{v_{\rm max}}
\def\bh{\bar{h}}
\def\del{\partial}
\def\nab{\nabla}
\def\half{{\textstyle{\frac{1}{2}}}}
\def\fourth{{\textstyle{\frac{1}{4}}}}

\def\bD{{\bf D}}
\def\bE{{\bf E}}
\def\bF{{\bf F}}
\def\bB{{\bf B}}
\def\bP{{\bf P}}
\def\bV{{\bf v}}
\def\bv{{\bf v}}
\def\bx{{\bf x}}
\def\by{{\bf y}}
\def\bz{{\bf z}}
\def\ba{{\bf a}}
\def\bd{{\bf d}}
\def\bs{{\bf s}}
\def\bn{{\bf n}}
\def\bp{{\bf p}}

\def\O{\Omega}

\def\br{{\bf r}}
\def\bnab{{\bf \nab}}

\def\tE{\tilde{E}}
\def\tL{\tilde{L}}
\def\Horava{Ho\v{r}ava }

\def\oxtwo{\mathscr{O}\left(x^2\right)}
\def\oxthree{\mathscr{O}\left(x^3\right)}
\def\oxfour{\mathscr{O}\left(x^4\right)}
\def\oxfive{\mathscr{O}\left(x^5\right)}

\def\ph{\phantom}
\def\LL{Lanczos-Lovelock}

\begin{document}
\title{The issue of zeroth law for Killing horizons in Lanczos-Lovelock gravity}
\author{Sudipta Sarkar}\email{sudiptas@imsc.res.in}\author{Swastik Bhattacharya}\email{swastik@imsc.res.in}
\affiliation{The Institute of Mathematical Sciences, Chennai, India}
\date{\today}
\begin{abstract}
We study the zeroth law for Killing horizons in Lanczos-Lovelock gravity. We show that the surface gravity of a general Killing horizon in Lanczos-Lovelock
gravity (except for general relativity) may not be constant even when the matter source satisfies dominant energy condition.
\end{abstract}
\maketitle
 General relativity (GR), being quantum mechanically non-renormalizable, may make sense as a Wilsonian effective theory working perturbatively in
powers of the dimensionless small parameter $G\,(Energy)^{D-2}$, where $G$ is the $D$-dimensional Newton's constant. Then the Einstein-Hilbert Lagrangian
is the lowest order term (other than the cosmological constant) in a derivative expansion of generally covariant actions for a metric theory, and the
presence of higher curvature terms is presumably inevitable. In general, the specific form of these terms will
 depend on the detailed features of the quantum gravity model. Still, from a purely classical point of view, a natural
 modification of the Einstein-Hilbert action is to include terms preserving the diffeomorphism invariance and still leading
 to an equation of motion containing no more than second order time derivatives. Interestingly, this generalization is unique
 \cite{Lanczos:1938sf, Lanczos-Lovelock:1971yv} and goes by the name of Lanczos-Lovelock gravity. Lanczos-Lovelock gravity is free
 from perturbative ghosts \cite{Zwiebach:1985uq} and leads to a well-defined initial value formalism \cite{ChoquetBruhat:1988dw}.
  The lowest order Lanczos-Lovelock correction term in space time dimensions $D > 4$, namely the Gauss-Bonnet term, also appears
   as a low energy $\alpha'$ correction in case of heterotic string theory \cite{Zwiebach:1985uq, Sen:2007qy}.
   Hence, it is interesting to pursue various classical and semi classical properties of Lanczos-Lovelock gravity.
   For example, the striking similarity of the laws of black hole mechanics with thermodynamics was
   first established in the case of general relativity \cite{Bardeen:1973gs} and a natural question is to ask whether this analogy is a peculiar property of
   GR or a robust feature of any generally covariant theory of gravity. Studying the properties of black holes in a general Lanczos-Lovelock theory may
   provide a partial answer to this important question.\\

The equilibrium state version of first law for black holes was established by Wald and collaborators
\cite{Wald:1993nt,Iyer:1994ys} for any arbitrary diffeomorphism invariant theory of gravity.
The entropy of the black hole can be expressed as a local geometric quantity integrated over a
space-like cross section of the horizon and is associated with the Noether charge of Killing isometry that generates the horizon.\\
It is also possible to write down a quasi stationary version of the second law for \LL\ gravity
\cite{Chatterjee:2011wj, Kolekar:2012tq} which proves that the entropy of black holes in \LL\
gravity monotonically increases for physical processes in which the horizon is perturbed by the
accretion of positive energy matter and the black hole ultimately settles down to a stationary state.\\

On the other hand, the zeroth law of black hole mechanics which asserts the constancy of
surface gravity has two independent versions. First, zeroth law can be established for stationary
 Killing horizons in GR, provided the matter obeys dominant energy condition  \cite{Bardeen:1973gs}.
  The other version states that the surface gravity is constant on the horizon of a static or
  stationary-axisymmetric black hole with the $t-\phi$ orthogonality property \cite{carter, Racz:1995nh}.
  The second version is entirely geometrical and independent of the field equation, whereas, the first
   version does not require the assumption of $t-\phi$ orthogonality property, but is only valid for GR,
   i.e. when Einstein's equation with matter obeying dominant energy condition holds.\\

Motivated by the fact that both the first law and a quasi stationary second law hold true for
\LL\ gravity, we study the zeroth law for a general Killing horizon in \LL\ gravity.
 We would like to see that if one uses \LL\ equation of motion and suitable energy condition,
  then whether it is possible to prove the constancy of surface gravity without any assumption of extra symmetry.
   In fact, in this paper, we provide a negative answer to this question and show that the constancy of
   surface gravity does not hold in general even when the matter source obeys dominant energy condition.\\

The paper is organized as follows: we first review the properties of Killing horizons. Next, we discuss the
Lanczos-Lovelock theory and present the main result. Finally, we conclude with further discussions. We adopt
the metric signature $(-, +, +, +, ...)$ and our sign conventions are same as those of \cite{Wald:1984rg}.\\
Also, we note that in general, a Killing horizon is not
necessarily an event horizon. But there is a version of rigidity
theorem \cite{carter} which states that for a static black hole,
the static Killing field must be normal to the horizon, whereas
for a stationary-axisymmetric black hole with the $t-\phi$
orthogonality property, there exists a Killing field which is
normal to the event horizon. In case of GR, it is possible to show
\cite{Hollands:2006rj, Hawking:1973uf} that the event horizon of a
stationary black hole is also a Killing horizon with no
assumptions of symmetries beyond stationarity. We are not aware of
any such proof for \LL\ gravity.

In a $D$-dimensional space time, a Killing horizon (not
necessarily an event horizon) is a null hyper-surface ${\cal H}$
whose null generators are the orbits of a Killing field $\xi^a =
(\partial/\partial v)^a$, which is null on the horizon. Then there
exists a function, surface gravity, ``$\kappa$'' of the Killing
horizon which is defined by the equation, \beq \xi^a \nabla_a
\xi^b = \kappa\, \xi^b. \label{def_kappa} \eeq
For static black holes, it is possible to provide a physical interpretation of the surface gravity.
In that case, surface gravity of the black hole horizon measures the force which must be exerted at
infinity to hold an unit mass at horizon. In general case, the surface gravity is the measure of the
failure of the Killing time to be the affine parameter along the horizon null generators. \\
From the definition in Eq.(\ref{def_kappa}), it is straightforward to show that the
surface gravity is constant along a generator \cite{Bardeen:1973gs,Wald:1984rg},
i.e. $\xi^a \nabla_a \kappa = 0$. In general, surface gravity may vary from one
generator to the other. Note that, the definition of surface gravity requires
the notion of stationarity. There is no notion of surface gravity for a general
non stationary dynamical horizon. \\
The real significance of the surface gravity is realized when one
consider quantum effects in a space time containing a black hole.
The semi classical calculations by Hawking \cite{Hawking:1974sw}
showed that the black hole emits thermal radiation with a
temperature (in units with $G = c = k = 1$), \beq T = \frac{\hbar
\kappa} { 2 \pi}. \eeq
Hawking's result immediately shows the importance of the zeroth law of black hole
mechanics as the zeroth law of black hole thermodynamics which states that the Hawking
temperature is uniform everywhere on a stationary black hole horizon.
This is reminiscent of the zeroth law of thermodynamics which states
that the temperature is uniform everywhere in a system in thermal equilibrium.\\
Since the constancy of the surface gravity along a generator of the horizon follows directly from the definition in Eq.(\ref{def_kappa}), we will only
discuss the change of ``$\kappa$'' from generator to generator. In order to proceed, we first construct a basis $\{\xi^a, N^a, e^{a}_A\}$ on the Killing
horizon where $\xi^a$ is the Killing field, $N^a$ is another null vector satisfying $\xi^a N_a = -1$, ${e^a}_{A}, \{ A = 2,...,D-1\}$ are the $(D-2)$ space
like vectors along the transverse directions and $\xi^b \gamma^{a}_{b} = N^b \gamma^{a}_{b} = 0$. Here $\gamma^{a}_{b}$ is the induced metric of any space
like slice of the horizon. We decompose the space time metric as $g_{ab} = g^{\perp}_{ab}\, +\, \gamma_{ab}$, where $g^{\perp}_{ab} = -2 \xi_{(a} N_{b)}$,
is the metric of the two dimensional space orthogonal to any horizon cross section. Also, for stationary space times with a Killing horizon, both the
expansion and shear vanish and using Raychaudhuri equation and the evolution equation for shear, we obtain \cite{Vega:2011ue,Wald:1984rg} that on the
horizon, \beq R_{ab} \xi^a \xi^b = 0 \label{stationary_conditions_1} \eeq and \beq \xi^a \gamma^{b}_{i}\gamma^{c}_{j} \gamma^{d}_{k} R_{abcd} = 0.
\label{stationary_conditions_2} \eeq We would like to emphasize that in order to derive these relationships, we have only used the fact that the horizon is
a Killing horizon with zero expansion and shear without assuming any further symmetry.

Next, we would like to study how the surface gravity changes from
one generator to the other. For that, we note that from
Eq.(\ref{def_kappa}), we can write $\kappa = - \xi^a N^b \nabla_a
\xi_b$ and then we obtain \cite{Bardeen:1973gs}, \beq
\gamma^{a}_{b} \nabla_a \kappa = - \xi^a R_{ac}
\gamma^{c}_{b}.\label{basic_eq} \eeq
So far, all the results are entirely geometrical and no use of the equation of motion is made. Now, if we further assume that the
horizon is  axisymmetric and possesses $t-\phi$ orthogonality property, then it is possible to show \cite{carter,Racz:1995nh}
that the RHS of Eq.(\ref{basic_eq}) vanishes identically and the surface gravity is constant on the horizon independent of the field equation. \\
Note that, if we assume that the Killing horizon possesses a bifurcation surface, i.e. a $(D-1)$ dimensional cross section on which the Killing field
$\xi^a$ vanishes, then $\gamma^{a}_{b} \nabla_a \kappa = 0$ on the bifurcation surface and since the surface gravity can not change along the generator,
this will establish the constancy of surface gravity on the entire Killing horizon. Hence, for Killing horizons with regular bifurcation surface, the
surface gravity is constant irrespective of gravitational dynamics \cite{Kay:1988mu}.
 But the assumption of the existence of bifurcation surface is a strong assumption and it is only applicable to a sub-class of Killing horizons.
 We would like to know if as in the case of general relativity, the constancy of surface gravity of Killing horizons in \LL\ gravity can be
 established without these assumptions.\\

 Let us now turn our attention to the features of \LL\ gravity. As discussed before, a natural generalization of
 the Einstein-Hilbert Lagrangian is provided by the Lanczos-Lovelock Lagrangian, which is the sum of dimensionally extended Euler densities,
\beq {\cal L}^{D} = \sum \limits_{m=0}^{[D-1)/2]} \alpha_m {\cal
L}_{m}^{D}, \eeq where the $\alpha_m$ are arbitrary constants and
${\cal L}_{m}^{D}$ is the $m$-th order Lanczos-Lovelock term given
by,
\begin{equation}
{\cal L}_{m}^{D}=\frac{1}{16\pi} \sum \limits_{m=0}^{[D-1)/2]}
\frac{1}{2^m} \delta^{a_1 b_1 \ldots a_m b_m}_{c_1 d_1 \ldots c_m
d_m} R^{c_1 d_1}_{~ a_1 b_1} \cdots R^{c_m d_m}_{~ a_m b_m},
\label{actionLL}
\end{equation}
where $R^{c d}_{~ a b}$ is the $D$ dimensional curvature tensor
and the generalized alternating tensor $\delta^{\ldots}_{\ldots}$
is totally anti-symmetric in both set of indices. For $ D = 2m$,
$16 \pi {\cal L}_{m}^{D}$ is the Euler density of $2m$-dimensional
manifold. The Einstein-Hilbert Lagrangian is a special case of
Eq.~(\ref{actionLL}) when $m=1$. The field equation of
Lanczos-Lovelock theory is, $G_{ab} + \alpha_m E_{(m)ab} = 8 \pi
T_{ab}$ where,
\begin{eqnarray}
E^{a}_{(m) b} = -  \frac{1}{2^{m+1}} \delta^{a a_1 b_1 \ldots a_m
b_m}_{b c_1 d_1 \ldots c_m d_m} R^{c_1 d_1}_{~ a_1 b_1} \cdots
R^{c_m d_m}_{~ a_m b_m},
\end{eqnarray}
 and $m \geq 2$. For convenience,  we have written the GR part (i.e. for $m = 1$)
 separately so that the GR limit can be easily verified by setting all $\alpha_{m}$'s to zero. \\
\LL\ gravity can be regarded as a natural and well behaved extension of general relativity
in higher dimensions. The spherically symmetric black hole solution in \LL\ gravity is derived
 in \cite{Boulware:1985wk,Boulware:1986dr}. The first law of black hole mechanics is studied
  in \cite{Jacobson:1993xs} and various thermodynamic properties of these black hole solutions
  are discussed in \cite{Myers:1988ze}. \\
Now, using the field equation of \LL\ gravity, we rewrite Eq.(\ref{basic_eq}) as, \beq \gamma^{a}_{b} \nabla_a \kappa = - 8 \pi \,\xi^a T_{ac}
\gamma^{c}_{b} + \alpha_m \xi_a E^{a}_{(m)c}\gamma^{c}_{b}.\label{form_kappa} \eeq

Next, to simplify the above expression, we would first like to show that for a  Killing horizon, $ E^{a}_{(m) b} \xi_a \xi^b = 0$. In order to prove that,
we first expand the curvature tensor in the basis $\{\xi^a, N^a, e^{a}_A\}$ on the horizon. Therefore, we write, \bea R^{c_1 d_1}_{a_1 b_1} =
g^{c_1}_{p}\,g^{d_1}_{q}\,g^{r}_{a_1}\,g^{s}_{b_1}\,R^{p q}_{r s}. \eea Now, as mentioned earlier, we express the space time metric as $g_{a b} =  -2
\xi_{(a} N_{b)} \, +\, \gamma_{ab}$. Also, stationarity ensures that some of the components are zero due to conditions Eq.(\ref{stationary_conditions_1})
and Eq.(\ref{stationary_conditions_2}). Then we can express,
\begin{widetext}
\begin{eqnarray}
 R^{c_1 d_1}_{a_1 b_1} &=&  \left(\gamma^{c_1}_p \gamma^{d_1}_q \gamma^r_{a_1} \gamma^s_{b_1}
                                          -2 \gamma^{c_1}_p\gamma^r_{a_1} \gamma^s_{b_1} N_q \xi^{d_1}
                                          + 4 \gamma^{c_1}_p \gamma^r_{a_1} \xi^{d_1} N_q \xi^s N_{b_1}
                                          +2 \gamma^r_{a_1} \gamma^{s}_{b_1} \xi^{c_1} N_p N^{d_1} \xi_q
                                          -4 \gamma^r_{a_1}  \xi^{c_1} N_p N^{d_1}\xi_q \xi^s N_{b_1} \right. \nonumber\\
                                           &-&\left. 2 \gamma^{c_1}_p \gamma^{d_1}_q   \gamma^r_{a_1} \xi_{b_1} N^s
                                          +2 \gamma^{c_1}_p \gamma^{d_1}_q \xi_{a_1} N^r \xi^s N_{b_1}
                                          +4 \gamma^{c_1}_p \gamma^r_{a_1} \xi^{d_1} \xi_{b_1} N_q N^s
                                             - 4 \gamma^{c_1}_p \xi^{d_1} N_q \xi_{a_1}  N^r \xi^s N_{b_1} \right. \nonumber\\
                                           &+& \left. 4 \gamma^{c_1}_p \gamma^r_{a_1} \xi_q N^{d_1} \xi_{b_1} N^s
                                           -  4 \gamma^{c_1}_p \xi_q N^{d_1}\xi_{a_1} N^r \xi^s N_{b_1}
                                          -4 \gamma^r_{a_1} \xi^{c_1}  N_p \xi_{q} N^{d_1} \xi_{b_1} N^s  \right. \nonumber\\
                                          &+& \left. 4 \xi^{c_1} N_p \xi_q N^{d_1} \xi_{a_1} N^r \xi^s N_{b_1}  \right)
                                                                               R^{pq}_{rs} \label{curvaturecomp}
\end{eqnarray}
\end{widetext}
We again remind the reader that this expression is valid only on the horizon. We also used the antisymmetry of the generalized alternating tensor
$\delta^{\ldots}_{\ldots}$. Now any component of a curvature tensor along the direction of the Killing vector in the expression of $ E^{a}_{(m) b}$  will
not contribute when contracted by $\xi_a \xi^b$. These constraints ensure that the only surviving contribution will be from the transverse components and
as a result we get, \beq E^{a}_{(m) c}\xi_a \xi^c &=& - \frac{1}{2^{m+1}} \delta^{a A_1 B_1 \ldots A_{m-1} B_{m-1}A_m B_m}_{c C_1 D_1 \ldots C_{m-1}
D_{m-1} C_m D_m} \nonumber \\ && R^{C_1 D_1}_{~ A_1 B_1}
 \cdots R^{C_m D_m }_{A_m B_m} \xi^c \xi_a,
\eeq where, indices $(A_1, B_1, C_1, ...)$ only take transverse values.\\

 Next, we explicitly break the alternating tensor as the totally antisymmetric
product of the Kronecker delta. For example, we write,
\begin{eqnarray}
&& \delta^{a \,a_1 b_1 \ldots a_{m-1} b_{m-1} a_m b_m}_{c \, c_1 d_1 \ldots c_{m-1} d_{m-1} c_m d_m } = \delta^{a}_{ c}\, \delta^{a_1 b_1 \ldots a_{m-1}
b_{m-1} a_m b_m}_{ c_1 d_1 \ldots c_{m-1} d_{m-1} c_m d_m } \nonumber \\&&- \,  \delta^{a}_{ c_1}\, \delta^{a_1 b_1 \ldots a_{m-1} b_{m-1} a_m b_m}_{ c\, d_1 \ldots c_{m-1}
d_{m-1 } c_m d_m } +  \delta^{a}_{ d_1}\, \delta^{a_1 b_1 \ldots a_{m-1} b_{m-1} a_m b_m}_{ c\, c_1 \ldots c_{m-1} d_{m-1} c_m d_m } \nonumber \\&&+ \,
...\label{expansion}
\end{eqnarray}
Note that, when contracted by $\xi_a \xi^c$,  the first term vanishes and also the rest of the terms do not contribute if all other indices are projected along
the transverse directions. 

Using this rule of expansion, we immediately see that on the horizon, $E^{a}_{(m) b} \xi_a \xi^b = 0$. Then,
Eq.(\ref{stationary_conditions_1}) and the field equation give that on the horizon $T^{a}_{b} \xi_a \xi^b = 0$. \\

Now, if the energy-momentum tensor obeys the dominant Energy Condition \cite{Wald:1984rg}, $T^{a}_{b} \xi^b $ will be a non-space like vector. But on the
horizon, we have seen that the field equation implies $T^{a}_{b} \xi_a \xi^b = 0$. Therefore, to obey dominant energy condition $T^{a}_{b} \xi^b $ must be
in the direction of $\xi^{a}$ only and as a result  $\xi^a T_{ac} \gamma^{c}_{b} = 0$. So, we ultimately arrive at, \beq \gamma^{a}_{b} \nabla_a \kappa =
\alpha_m \xi_a E^{a}_{(m)c}\gamma^{c}_{b}.\label{form_kappa_final} \eeq
From the above equation, setting $\alpha_m = 0$, we can obtain the result of \cite{Bardeen:1973gs}, which proves the constancy of surface gravity for GR.\\

We will now show that on the Killing horizon, $ \xi^a
E_{(m)ac}\gamma^{c}_{b}$ does not vanish identically unless one
imposes additional constraints on the geometry of the horizon. \\

To prove this, we again consider the expansion of the curvature
tensor $R^{c_1 d_1}_{a_1 b_1}$ in the basis $\{\xi^a, N^a,
e^{a}_A\}$ on the horizon and use
Eq.(\ref{stationary_conditions_1}) and
Eq.(\ref{stationary_conditions_2}). Due to the antisymmetry of the
generalized alternating tensor $\delta^{\ldots}_{\ldots}$, any
component along $\xi_{a_1}$ or $\xi_{b_1}$ in the expression
Eq.(\ref{curvaturecomp}), will not contribute when contracted by
$\xi_a$. Then, the only non zero contributions of the expansion of
$R^{c_1 d_1}_{a_1 b_1}$ in the basis $\{\xi^a, N^a, e^{a}_A\}$ are,
\begin{widetext}
\beq && R^{C_1 D_1}_{A_1 B_1} - 2 N_p R^{C_1 p}_{A_1 B_1}
\xi^{d_1} + 4 N_p \xi^q R^{C_1 p}_{A_1 q} \xi^{d_1} N_{b_1} + 2
N_p \xi_q R^{p q}_{A_1 B_1} \xi^{c_1} N^{d_1} - 4 N_p \xi_q \xi^r
R^{p q}_{A_1 r}  \xi^{c_1} N^{d_1} N_{b_1}. \eeq
\end{widetext}
We now consider products of two curvatures of the form $R^{c_{m-1}
d_{m-1}}_{a_{m-1} b_{m-1}}R^{c_m d_m}_{a_m b_m}$. Due to the
antisymmetry of the generalized alternating tensor
$\delta^{\ldots}_{\ldots}$, the products of the components along
the direction of the Killing vector will not contribute and the
non vanishing contributions in the product $R^{c_{m-1}
d_{m-1}}_{a_{m-1} b_{m-1}}R^{c_m d_m}_{a_m b_m}$ can be expressed
as,
\begin{widetext}
\beq &&
 \,R^{C_{m-1} D_{m-1}}_{A_{m-1} B_{m-1} } \left[ R^{C_m D_m}_{A_m B_m} - 4 N_p R^{C_m p}_{A_m B_m} \xi^{d_m} + 8 N_p \xi^q R^{C_m p}_{A_m q}
 \xi^{d_m} N_{b_m} + 4 N_p \xi_q R^{p q}_{A_m B_m} \xi^{c_m} N^{d_m} \right. \nonumber \\ && \left.  - 8 N_p \xi_q \xi^r R^{p q}_{A_m r}
 \xi^{c_m} N^{d_m} N_{b_m} \right] \label{final_componets}
\eeq
\end{widetext}
Continuing in this way, we can express the product of
$m$-curvature tensors appearing in $ \xi^a
E_{(m)ac}\gamma^{c}_{b}$ as,

\begin{widetext}
\beq &&
 R^{C_1 D_1}_{A_1 B_1} \dots R^{C_{m-1} D_{m-1}}_{A_{m-1} B_{m-1} } \left[ R^{C_m D_m}_{A_m B_m} -
 2^{m} \left(  N_p R^{C_m p}_{A_m B_m} \xi^{d_m} - 2 N_p \xi^q R^{C_m p}_{A_m q} \xi^{d_m} N_{b_m} -
 N_p \xi_q R^{p q}_{A_m B_m} \xi^{c_m} N^{d_m} \right. \right. \nonumber \\ && \left. \left.  +
 2 N_p \xi_q \xi^r R^{p q}_{A_m r}  \xi^{c_m} N^{d_m} N_{b_m} \right) \right]. \label{final_componets1}
\eeq
\end{widetext}
This entire expression is contracted with the alternating tensor $\xi^a\gamma^{c}_{b} \delta^{\ldots}_{\ldots}$. Again, the expansion of the alternating
tensor in Eq. (\ref{expansion}) ensures that the first term in the above expansion is zero when contracted by $\xi_a \gamma^{c}_{b}$. Also, using the expansion rule in Eq.(\ref{expansion}), it is straightforward to see that the only non-vanishing contribution comes from the last term in Eq.(\ref{final_componets1}), given by,
\begin{widetext}
\begin{eqnarray}
 && \delta^{a\, a_1 b_1 \ldots a_{m-1} b_{m-1} a_m b_m}_{c\, c_1 d_1 \ldots c_{m-1} d_{m-1} c_m d_m }R^{C_1 D_1}_{A_1 B_1} \dots R^{c_{m-1} d_{m-1}}_{a_{m-1} b_{m-1}}
 R^{c_m d_m}_{a_m b_m}\xi_a \gamma^{c}_{b}  \nonumber \\
&&= \xi_a \gamma^{c}_{b} \delta^{a\, A_1 B_1 \ldots A_{m-1} B_{m-1} A_m b_m}_{c \,C_1 D_1 \ldots C_{m-1} D_{m-1} c_m d_m }
 R^{C_1 D_1}_{A_1 B_1} \dots R^{C_{m-1} D_{m-1}}_{A_{m-1} B_{m-1} } R^{p q}_{A_m r}  N_p \,\xi_q\, \xi^r \,\xi^{c_m} N^{d_m} N_{b_m}.  \nonumber \\
\label{dawood}
\end{eqnarray}
\end{widetext}
Again, using the rules of expansion for the alternating tensor, it is straightforward to show that the only non zero contribution from this term is of the
form, 
\begin{widetext}
\begin{eqnarray} && \xi_a \gamma^{c}_{b} \delta^{a}_{d_m} \, \delta^{b_m}_{c_m}\delta^{ A_1 B_1 \ldots A_{m-1} B_{m-1} A_m }_{c\, C_1 D_1 \ldots
C_{m-1} D_{m-1} }
 R^{C_1 D_1}_{A_1 B_1} \dots R^{C_{m-1} D_{m-1}}_{A_{m-1} B_{m-1} } R^{p q}_{A_m r}  N_p \, \xi_q \,\xi^r \, \xi^{c_m} N^{d_m} N_{b_m} \nonumber \\
 &&= 2^{m+1}\,\delta^{ A_1 B_1 \ldots A_{m-1} B_{m-1} A_m}_{B \, C_1 D_1 \ldots C_{m-1} D_{m-1} }R^{C_1 D_1}_{A_1 B_1} \dots R^{C_{m-1} D_{m-1}}_{A_{m-1}
B_{m-1} } R^{p \,q}_{A_m r}N_p\, \xi_q\, \xi^r. 
\end{eqnarray}
\end{widetext}

Since, for stationary black holes, both the expansion and shear vanish, we can write \cite{Wald:1984rg}, $ R^{C D}_{A B} =^{(D-2)}R^{C D}_{A B}$,
where, $^{(D-2)}R^{C D}_{A B}$ is the intrinsic curvature of the cross section of the horizon. Then, we recall the expression for the equation of
motion of a $(m-1)$-th order \LL\ theory constructed using intrinsic curvatures of the horizon cross section, which is given by,
\begin{widetext}
\begin{eqnarray} && \ ^{(D-2)}E^{a}_{(m-1)b} = \ -  \frac{1}{2^{m}} \delta^{a\, a_1 b_1 \ldots a_{m-1} b_{m-1} }_{b\, c_1 d_1 \ldots c_{m-1} d_{m-1}} \, ^{(D-2)}R^{c_1
d_1}_{~ a_1 b_1} \cdots ^{(D-2)}R^{c_{m-1} d_{m-1}}_{~ a_{m-1} b_{m-1}}. 
\end{eqnarray}
\end{widetext}
Using this expression, we finally arrive at, 
\beq
\left(2^{-m}\right) \gamma^{a}_{b} \nabla_a \kappa = -\alpha_m \, ^{(D-2)}E^{a}_{(m-1)b} R^{p\, q}_{a\, r}\,N_p \xi_q \xi^r, \label{last_exp}
 \eeq
 Eq. (\ref{last_exp}) is our final expression and the right hand side of this equation does not vanish in general. So, unlike general relativity, for any
 higher order \LL\ theory of gravity, the surface gravity may vary from one generator to another on the horizon. As a result, although the surface
 gravity is constant along a single generator (i.e. surface gravity is independent of the affine parameter $\lambda$ of the horizon),
 it may depend on the angular coordinates and vary as one moves across the generators. \\

In case of general relativity, the constancy of surface gravity for Killing horizons can be proved without any other assumption. Then, the rigidity theorems \cite{Hawking:1973uf, Hollands:2006rj} ensure that every stationary event horizon in GR must be a Killing horizon and this in turn, proves the constancy of the surface gravity for stationary black holes. Our work shows that higher order \LL\ terms do not share this property and if there exits a stationary solution of \LL\ gravity with Killing horizon, which is not axisymmetric with $t-\phi$ symmetry, the surface gravity in
general will be a function of the angular coordinates on a cross section of the horizon. \\

Let us now consider the special cases. First of all, if the cross section of the horizon is flat, i.e when the horizon topology is planer,
 all the intrinsic curvature tensors of the horizon cross section vanish and as a result,
  the surface gravity will not vary from one generator to another. \\

  Also, if $R^{p\, q}_{b\, r}\,N_p\, \xi_q\, \xi^r = 0$ on the horizon, the surface gravity is
  again constant. A possible way to achieve this is to consider a stationary
  axisymmetric horizon with $t-\phi$ isometry. For example, using equation $2.18$ and $2.27$ in \cite{Vega:2011ue},
   we can show that if the expansion and shear vanish, as in the case of a stationary black hole, then we have, $R^{p\, q}_{B\, r}\,N_p\, \xi_q\,
   \xi^r = \xi^a R_{ac} \gamma^{c}_{b}$, for a stationary horizon.
   Now, a sufficient condition which ensures that $ \xi^a R_{ac} \gamma^{c}_{b}= 0$, is the existence of a stationary and axisymmetric horizon
   with $t-\phi$ isometry \cite{Wald:1984rg}. This is basically the result obtained in \cite{Racz:1995nh}. \\

Also, so far, we do not have any stationary solution of \LL\ gravity except general relativity. But, at least for general relativity with Gauss Bonnet
correction terms, we have explicit spherically symmetric solutions \cite{Boulware:1985wk, Boulware:1986dr}. Given the quasi-linearity of the field
equations of \LL\ gravity, it is quite possible that in this theory, a stationary solution will be found. If such solutions exist, our work shows that these
black hole solutions will have non-constant surface gravity unless they are axisymmetric with $t-\phi$ symmetry. Also, once we consider quantum effects,
the surface gravity is proportional to the Hawking temperature of the black hole and hence, if the surface gravity is no longer constant on the horizon and
varies from one generator to another, then we can not treat such a stationary black hole as a system in thermodynamic equilibrium. \\

But, there is still a possibility that as in the case of general relativity, all stationary horizons in a \LL\ theory are axisymmetric with $t-\phi$
isometry. If that happens, then the zeroth law will be valid automatically. In order to investigate this, one needs to try for a generalization of the
strong rigidity theorem \cite{Hawking:1973uf, Hollands:2006rj} for \LL\ gravity. The proof of the rigidity theorem depends on the initial value formulation
of Einstein's equations. Since, the field equations of \LL\ gravity are also second order in time, and as a result the initial value formalism is well
defined, it is reasonable to expect the validity of rigidity theorem for \LL\ gravity.\\

Another obvious generalization of our work will be to study the zeroth law of Killing
horizons in a general diffeomorphism invariant theory of gravity. Although, the
techniques used in this work are to some extent specific to only \LL\ gravity,
still we can provide some requirements which will ensure the validity of the zeroth law for any diffeomorphism invariant theory.\\
 To begin with, let us consider a general diffeomorphism invariant theory of gravity described
 by an Lagrangian ${\cal L}$. Suppose, the field equation of the theory is given by
 $ {\cal E}_{ab} = 8 \pi T_{ab}$, where ${\cal E}_{ab}$ represents a covariantly conserved, symmetric tensor obtained from the variation of the
Lagrangian ${\cal L}$. Then, from Eq.(\ref{basic_eq}), we obtain, \beq
\gamma^{a}_{b} \nabla_a \kappa &=& - \xi^a R_{ac} \gamma^{c}_{b}\nonumber \\
&=&  \xi^a \left( {\cal E}_{ac} - R_{ac}\right) \gamma^{c}_{b} - 8
\pi \, \xi^a T_{ac} \gamma^{c}_{b}. \eeq The zeroth law will hold
if on the Killing horizon, following constraints are satisfied,
\beq
 {\cal E}_{ab}  \xi^a \xi^b = 0 ~~~ \textrm{and}~~~\left( {\cal E}^{a}_{c} - R^{a}_{c}\right) \xi_a \gamma^{c}_{b} = 0
\eeq
In general, it is difficult to check these constraints for a general gravity theory,
but if the above two conditions hold and the matter source obeys dominant energy then
that will be enough to ensure the constancy of surface gravity on the Killing horizon
of a stationary space time. In case of general relativity, both of these constraints
are satisfied and the zeroth law holds true even for a general Killing horizon. For
\LL\ theory, the first constraint holds, but the second one is not true in general
and as a result, the surface gravity is no more constant on the horizon.\\
 In fact, it is also quite possible that the zeroth law does not hold for a general
  stationary black hole in some class of gravity theories. In that case, this may
  be useful as a criterion to select a sub class of diffeomorphism invariant actions
   as preferred theories where a consistent formulation of black hole thermodynamics is possible.

\section*{Acknowledgments}
We are especially grateful to Ted Jacobson for detailed comments on a previous draft of this article. We would like to thank Ghanashyam Date and T
Padmanabhan for comments and discussions. We also thank the anonymous referee for various helpful suggestions to improve the presentation of the results.

\end{document}